\documentclass[12pt,english]{article}%
\usepackage{amsmath}
\usepackage{amsfonts}
\usepackage{amssymb}
\usepackage{graphicx}%
\newcommand{\be}{\begin{equation}}
\newcommand{\ee}{\end{equation}}
\newcommand{\bea}{\begin{eqnarray}}
\newcommand{\eea}{\end{eqnarray}}
\newcommand{\ba}{\begin{array}}
\newcommand{\ea}{\end{array}}
\newcommand{\p}[1]{(\ref{#1})}

\def\bbox{{\,\lower0.9pt\vbox{\hrule \hbox{\vrule height 0.2 cm
\hskip 0.2 cm \vrule height 0.2 cm}\hrule}\,}}
\newcommand{\dsl}{\pa \kern-0.5em /}

\font\mybb=msbm10 at 10pt
\def\bb#1{\hbox{\mybb#1}}

\def\bC {\bb{C}}

\makeatother
\begin{document}



\begin{titlepage}
\rightline{\tt{hep-th/0612300}}
\rightline{UMTG-4, JINR-E2-2006-150, DAMTP-06-107}
\vfill
\begin{center}
\baselineskip=16pt {\Large  {\bf Planar Super-Landau Models Revisited}}

\vskip 0.3cm
{\large {\sl }}
\vskip 10.mm
{Thomas Curtright$^{\dagger,1}$, ~Evgeny Ivanov$^{*,2}$, Luca Mezincescu$^{\dagger,3}$,\\
and Paul K. Townsend$^{+,4}$}
\vskip 1cm
{\small
$^\dagger$
Department of Physics, University of Miami,\\
Coral Gables, FL 33124, USA\\
}
\vspace{6pt}
{\small
$^*$
Bogoliubov Laboratory of Theoretical Physics, JINR\\141980 Dubna, Moscow Region,
Russia\\
}
\vspace{6pt}
{\small
$^+$
Department of Applied Mathematics and
Theoretical Physics\\
Centre for Mathematical Sciences, University of
Cambridge\\
Wilberforce Road, Cambridge, CB3 0WA, UK\\
}
\end{center}%
\vfill
\par
\begin{center}
{\bf ABSTRACT}
\end{center}
\begin{quote}
We use the methods of  ${\cal PT}$-symmetric quantum theory to find a one-parameter
family of $ISU(1|1)$-invariant planar super-Landau models with positive norm,
uncovering an `accidental', and generically spontaneously-broken,
worldline supersymmetry, with charges that have  a Sugawara-type representation
in terms of the $ISU(1|1)$ charges.
In contrast to standard models of supersymmetric
quantum mechanics, it is the norms of states rather than their energies
that are parameter-dependent, and the spectrum changes discontinuously
in the limit that worldline supersymmetry is restored.

\vfill
\vfill
\vfill
\hrule width 5.cm
\vskip
2.mm
{\small
\noindent $^1$  curtright@physics.miami.edu\\
\noindent $^2$
eivanov@theor.jinr.ru\\
\noindent $^3$ mezincescu@server.physics.miami.edu%
\\
\noindent $^4$ p.k.townsend@damtp.cam.ac.uk
\\ }
\end{quote}
\end{titlepage}

\section{Introduction}

\setcounter{equation}{0}

In recent works, three of us have explored the mathematics and physics of `super-Landau' models,
which are quantum mechanical models for a charged particle on a homogeneous superspace,
such that the `bosonic' truncation is either Landau's original model for a charged
particle moving on a plane under the influence of a uniform magnetic field,
or Haldane's spherical version of it. The former case yields `planar' super-Landau
models and the latter case yields `spherical' super-Landau models; the two
are related by a limiting process in which the sphere becomes a plane
as its radius is taken to infinity.

The simplest spherical super-Landau models are those for which the homogeneous
superspace has isometry supergroup $SU(2|1)$, and the simplest of these  is
the `superspherical' Landau model for a charge particle on the projective
superspace $CP^{(1|1)}$, which is the complex `Riemann supersphere'
and can be viewed as the coset superspace $SU(2|1)/U(1|1)$.   In a limit
in which only the lowest Landau level (LLL) of this model survives,
it describes a fuzzy Riemann supersphere \cite{Ivanov:2003qq}. The other
spherical super-Landau models with  $SU(2|1)$ symmetry are the
 `superflag' Landau models, for which the homogeneous superspace is the coset superspace
$SU(2|1)/[U(1)\times U(1)]$ \cite{Ivanov:2004yw}. In this case there is
an additional anticommuting variable,  and a corresponding `fermionic Wess-Zumino'  term
with real number  coefficient  $M$. There is therefore a 1-parameter family of superflag Landau models,
and the $M{=}0$ model turns out to be equivalent to the superspherical model.

The quantum theory of the spherical super-Landau models  was worked out
in \cite{Ivanov:2003qq,Ivanov:2004yw} and a number of intriguing properties were uncovered.
The spherical models are conceptually simpler than the planar models because
the degeneracies at each Landau level are finite, but  the non-linearity
of the configuration space leads to computational complexities.
For this reason, it is useful to study the class of planar super-Landau models
obtained as the planar limit of the spherical super-Landau models; these
all have isometry supergroup $ISU(1|1)$. The planar limit of the superspherical model
yields the `superplane' Landau model, while the planar limit of the superflag Landau
models yields the  `planar-superflag' Landau models \cite{Ivanov:2005vh},
which are parametrized by the real number $M$, with the $M{=}0$ model being
equivalent to the  superplane model.

One result of \cite{Ivanov:2003qq,Ivanov:2004yw,Ivanov:2005vh} was
that there are ghosts in all Landau levels with $N{>}2M$, and zero-norm states
in all levels with $N{=}2M$ (which is possible when $2M$ is a non-negative integer).
This result assumes a natural superspace norm, invariant under the superspace
isometries and with respect to which the  Hamiltonian is hermitian, and it  shows that
this norm is indefinite. This was not unexpected since the classical equations
of motion for the `fermionic'  variables are (except in the LLL limit) second
order in time derivatives, rather than first order; this typically leads to ghosts in quantum
field theory, and in quantum  mechanics \cite{Volkov:1980mg}.  However, more options
are available in quantum mechanics\footnote{A recent article \cite{Robert:2006kp} shows
that even  `bosonic'  ghosts need not lead to a violation of unitarity in quantum
mechanics.}. In particular,  the possibility of an alternative norm
was not addressed in \cite{Ivanov:2003qq,Ivanov:2004yw,Ivanov:2005vh},
although it is not difficult to see that there must exist a positive norm:
the hermiticity of the Hamiltonian with respect to any non-degenerate
norm implies that it is both diagonalizable and  has real eigenvalues,
and it is therefore manifestly hermitian with respect to the natural
positive-definite norm in the basis in which it is diagonal. However, it is not immediately
clear what the consequences are for the symmetries, nor whether there
are further possibilities. One purpose of this paper is  to explore the possibilities
for symmetry-preserving norms that maintain the hermiticity of the Hamiltonian,
and  thereby to determine whether the ghosts found previously in super-Landau
models can be `exorcized'.

In order to simplify the calculations we will restrict ourselves
here to the planar super-Landau models.
We address the issue of the uniqueness, or otherwise, of the Hilbert space
norm by adapting the methods of  ${\cal PT}$-symmetric quantum theory
(see \cite{BenderReview} for a review) and  `bi-orthogonal systems'
(see e.g. \cite{Curtright:2005zk}). In that context one is given
a Hamiltonian that fails to be hermitian with respect to  a `naive'
norm and one considers whether it is possible to deform the norm
in such a way  that the Hamiltonian becomes hermitian. In our case,
the starting Hamiltonian is hermitian but with respect to an
indefinite norm and we need to modify the norm so that it becomes
positive and is such that the Hamiltonian remains hermitian. The
two problems are quite different but, in either case,  the formalism
is well-adapted to study the consequences for symmetries of a change of norm.
Our conclusion will be that  for the planar superflag there are {\it two}
possible $ISU(1|1)$ invariant norms, one being the norm used
in \cite{Ivanov:2003qq,Ivanov:2004yw,Ivanov:2005vh}. For $M{<}0$, it is
the `other' possible norm that is both positive and $ISU(1|1)$ invariant,
but zero-norm states appear for $M{=}0$ and positivity for $M{>}0$ requires
a `dynamical' combination of both possible norms.

The issue of the Hilbert space norm was indirectly brought to our attention
by a paper of Hasebe \cite{Hasebe:2005cm} on an alternative `superplane'
Landau model obtained as the planar limit of a spherical super-Landau model
for a particle on the coset superspace  $OSp(1|2)/U(1)$  \cite{Hasebe:2004hy}.
A feature of $OSp(1|2)/U(1)$, which is also referred to as a `supersphere'
by many authors,  is that the `fermions'  transform as an $SU(2)$ doublet,
which means that they must be complex because the doublet of $SU(2)$
is pseudo-real rather than real. This feature carries over to the planar limit,
so the `superplane' of Hasebe is a superspace of real dimension $(2|4)$
in contrast to the $(2|2)$-dimensional superplane of  \cite{Ivanov:2005vh},
but  it  can be interpreted as a superspace of `pseudo-real'
dimension $(2|2)$ and it appears that the distinction is not relevant
to the quantum theory. A further difference between \cite{Ivanov:2005vh}
and \cite{Hasebe:2005cm}  is that  wave-functions were interpreted
in  \cite{Ivanov:2005vh} as superfields (functions of definite Grassmann parity),
and this leads to a `Hilbert' supervector space rather than to a standard vector
space.  In contrast, the coefficients in the $\zeta$-expansion of the wave-functions 
in \cite{Hasebe:2005cm}  are all standard complex functions, and the norm for the
Hilbert space  they span is the positive-definite one. A remarkable
feature of this choice is that the quantum theory can then be interpreted
as a model of supersymmetric quantum mechanics (SQM); specifically, it has
an unbroken ${\cal N}{=}2$ `worldline' supersymmetry\footnote{Here we adopt
the convention that ${\cal N}$ is the number of  {\it real} supercharges, in contrast to
\cite{Hasebe:2005cm, Hasebe:2004hy}, and  \cite{Witten:1981nf},
where ${\cal N}$ is the number of  {\it complex} supercharges.},  the SQM ground states forming the lowest Landau level \cite{Hasebe:2005cm}.

The emergence of worldline supersymmetry is remarkable because it has no obvious
connection to the `internal'  supersymmetry that underlies the model's construction,
and a major purpose of this paper is to elucidate its origin. As we shall see,
the transition from the indefinite norm to the positive-definite one leads to a change
in the conjugation properties of `fermionic' operators, which is effected via
a `shift' operator, and this leads to the Hamiltonian appearing as a central charge
in the $ISU(1|1)$ algebra. In addition, the shift operator turns out to be the
supersymmetry charge of the worldline supersymmetry algebra.  A remarkable feature
of the worldline supersymmetry generators is that they have a Sugawara-type realization
in terms  of the original $ISU(1|1)$ generators, although this feature is absent in a
new `natural' basis for which the symmetry algebra  is manifestly the direct sum of the Lie
superalgebra of $ISU(1|1)$ and the ${\cal N}{=}2$ worldline supersymmetry superalgebra.

The status of worldline supersymmetry is considerably clarified by consideration of the
planar  superflag Landau models.  The additional anticommuting variable of these models
was identified in \cite{Ivanov:2005vh} as a Nambu-Goldstone variable for the $ISU(1|1)$ supersymmetry. However, this variable is actually $ISU(1|1)$-inert  in the `natural' basis, and instead transforms inhomogeneously under worldline supersymmetry (at least for $M\leq 0$, which we assume 
for the  purposes of this introduction); it is therefore a Nambu-Goldstone variable
for a spontaneously broken ${\cal N}{=}2$ worldline supersymmetry!   In the quantum
theory, this new anticommuting variable becomes the complex Grassmann-odd coordinate
of  worldline superspace, and the wavefunction becomes a worldline superfield,
with an expansion in terms of $ISU(1|1)$  superfields.

The equivalence of the $M{=}0$ planar superflag model to the superplane model was proved in
\cite{Ivanov:2005vh} for the indefinite norm, but we show here that it remains true for the new, positive,
norm. This equivalence means that the worldline supersymmetry that is spontaneously
broken for $M{<}0$ is restored when $M{=}0$. Classically, this is because $\xi$ becomes a pure-gauge variable
in the classical ground state when $M=0$.  Quantum mechanically, the worldline supersymmetry
restoration occurs because  half the ground states have zero norm when  $M{=}0$,
and the physical ground states (defined as equivalence classes of states modulo
the addition of a zero-norm state) are annihilated by the worldline supersymmetry
operators. In other words, supersymmetry is restored at $M{=}0$ by virtue
of a discontinuity  in the spectrum at $M{=}0$.  This is rather different
from the usual state of affairs for a family of SQM models in which
the spectrum changes continuously with the parameter, so that  supersymmetry
can be broken at some values of the parameter only if the Witten index
vanishes \cite{Witten:1981nf}. Here, it is not the energy eigenvalues
that depend on the parameter but the norms of the states, and this allows
a discontinuity  in the spectrum because the norms of some ground states can go to zero.

\section{Preliminaries}
\label{sec:prelim}
\setcounter{equation}{0}

It is useful to discuss first some of the general structures to be encountered
later in specific models. The quantum systems of interest possess inner
products which, while naturally defined, are not positive definite.
Therefore, let us assume that there exists a complete system of energy
eigenvectors $\left\vert f_{A}\right\rangle $ for the Hamiltonian, $H$, which
obey
\begin{equation}
\label{equation}\left\langle f_{A}|f_{B}\right\rangle =\left(  -\right)
^{g\left(  A\right)  }\delta_{AB}\,,
\end{equation}
where $g\left(  A\right)  $ is the grading\footnote{This should not be confused
with the grading associated to Grassmann parity, with anticommuting variables
being Grassmann-odd.}
\begin{equation}
g\left(  A\right)  =\left\{
\begin{array}
[c]{r@{\quad: \quad}lc}
0 & A=a & \cr1 & A=\alpha
\end{array}
\right..
\end{equation}
The subset of indices $A=a$ indicates positive norm states, while the subset
$A=\alpha$ indicates negative norm states, for all eigenvectors. In fact
(\ref{equation}) defines a system of linear functionals
\begin{equation}
\mathcal{F}_{A}\left(  f_{B}\right)  = \left(  -\right)  ^{g\left(  A\right)
}\delta_{AB}\,,
\end{equation}
which upon a trivial redefinition can be cast in the standard biorthogonal
form \cite{Curtright:2005zk} (see also \cite {Goursat, Banach,  dieu}).

The operation of \emph{naive} hermitian conjugation $\left(  \dagger\right) $
will be taken with respect to the non-positive-definite inner product. For
all the models of interest here, $H$ will be naively hermitian with respect to
this inner product:
\begin{equation}
H=H^{\dagger}\,.
\end{equation}
To define an improved inner product, and obtain only positive norms, we
introduce a `metric operator'  $G$ that acts
on the eigenvectors $\left\vert f_{A}\right\rangle $ to give
\begin{equation}
G\left\vert f_{A}\right\rangle \equiv\left\vert G f_{A}\right\rangle = \left(
-\right)  ^{g\left(  A\right)  }\left\vert f_{A}\right\rangle,\qquad
G=G^{\dagger}\, . \label{metric}
\end{equation}
Thus, $H$ commutes with the metric, essentially by definition of the grading.
The new inner product is then defined by the following formula
\begin{equation}
\left\langle \left\langle f_{A}|f_{B}\right\rangle \right\rangle
\equiv\left\langle Gf_{A}|f_{B}\right\rangle = \delta_{AB}\, . \label{inner}
\end{equation}
The `improved' hermitian conjugate $\mathcal{O}^{\ddagger}$, with respect
to $\left\langle \left\langle \cdots\right\rangle \right\rangle$,
of any operator $\mathcal{O}$,  is given by
\begin{equation}
\left\langle Gf_{A}\left\vert \mathcal{O}\right\vert f_{B}\right\rangle
=\left\langle \mathcal{O}^{\dagger}Gf_{A}|f_{B}\right\rangle =\left\langle
G\left(  G^{-1}\mathcal{O}^{\dagger}G\right)  f_{A}|f_{B}\right\rangle.
\end{equation}
That is to say,
\begin{equation}
\left\langle \left\langle f_{A}\left\vert \mathcal{O}\right\vert
f_{B}\right\rangle \right\rangle =\left\langle \left\langle \mathcal{O}
^{\ddagger}f_{A}|f_{B}\right\rangle \right\rangle,
\end{equation}
where
\begin{equation}
\mathcal{O}^{\ddagger}\equiv G^{-1}\mathcal{O}^{\dagger}G=\mathcal{O}
^{\dagger}+S_{\mathcal{O}}\, . \label{shift1}
\end{equation}
Here we have introduced a \textquotedblleft shift operator\textquotedblright
\ for a given $\mathcal{O}$,\ as defined by
\begin{equation}
S_{\mathcal{O}}\equiv G^{-1}\left[  \mathcal{O}^{\dagger},G\right].
\label{shift2}
\end{equation}
Operators which do not commute with $G$ will have $\mathcal{O}^{\ddagger}
\neq\mathcal{O}^{\dagger}$.

Note that $G=G^{\dagger}$ implies $\left(  \mathcal{O}^{\ddagger}\right)
^{\ddagger}=\mathcal{O}\,$, so that the new hermitian conjugation procedure closes in
the familiar way. Correspondingly, the shift operators have the simple, useful
conjugation property
\begin{equation}
S_{\mathcal{O}}{}^{\ddagger}=-S_{\mathcal{O}}{}^{\dagger}\, . \label{shift3}
\end{equation}
As a consequence, the combination

\begin{equation}
\tilde{\mathcal{O} }\equiv \mathcal{O}+ \frac{1}{2}S_{\mathcal{O}}^{\dagger}
\label{O1}
\end{equation}
has a conjugation with respect to the metric that coincides with its naive
hermitian conjugate
\begin{equation}
\tilde{\mathcal{O}}^{\ddagger}=\tilde{\mathcal{O}}^{\dagger}\, . \label{shift4}
\end{equation}
We are going to extensively use all of these properties, as well as the
following proposition \newline

\noindent\textbf{[Lemma]} \ Since $\left[  G,H\right]  =0\,$, the Hamiltonian
$H$ is hermitian in both inner products, $H=H^{\dagger}=H^{\ddagger}\,$.
Moreover, if the operator $\mathcal{O}$ is a constant of motion, then the
corresponding shift operator is also a constant of motion.  Indeed, from
$\left[  \mathcal{O},H\right]  =0$ it follows that $\left[  \mathcal{O}
^{\dagger},H\right]  =0$ \emph{and} $\left[  \mathcal{O}^{\ddagger},H\right]
=0\,$. This is a signal that the algebra of operators which are in involution
with the Hamiltonian may be larger than originally assumed:  the system may
have some `hidden' symmetries.

\section{Fermionic Landau model}

\setcounter{equation}{0}

The fermionic Landau model \cite{Hasebe:2005cm,Ivanov:2005vh} has the Lagrangian
\be
L_f=\dot\zeta\dot{\bar\zeta} - i\kappa\left(\dot\zeta \bar\zeta
+ \dot{\bar\zeta}\zeta\right), \label{FermLL}
\ee
where $\kappa$ is a real positive constant, $\zeta(t)$ is a complex anticommuting variable
with complex conjugate $\bar\zeta(t)$, and the overdot indicates its derivative
with respect to the time parameter $t$. The equivalent phase space Lagrangian is
\begin{equation}
\tilde L_{f}=-i\dot{\zeta}\pi-i\dot{\bar{\zeta}}\bar{\pi}-H_{f}\, ,\qquad H_{f}=\left(
\bar{\pi}-\kappa\zeta\right)  \left(  \pi-\kappa\bar{\zeta}\right),
\end{equation}
where $\pi$ ($\bar{\pi}$) is the momentum conjugate to $\zeta$ ($\bar{\zeta}$).
To quantize, we make the replacements
\begin{equation}
\pi\rightarrow\partial_{\zeta}\, ,\qquad  \bar{\pi}\rightarrow\partial
_{\bar{\zeta}}\,, \label{replaceferm}
\end{equation}
where the Grassmann-odd derivatives should be understood as left derivatives.
With a standard operator ordering prescription, the quantum Hamiltonian is
\begin{equation}
H_{f}=\frac{1}{2}\left[\alpha, \alpha^\dagger\right] =
-\alpha^{\dagger}\alpha-\kappa\,, \label{fermham}
\end{equation}
where
\begin{equation}\label{fermiosc}
\alpha=\left(  \partial_{\bar{\zeta}}-\kappa\zeta\right),\qquad \alpha
^{\dagger}=\left( \partial_{\zeta}-\kappa\bar{\zeta}\right).
\end{equation}
These operators satisfy the anticommutation relations
\begin{equation}
\left\{ \alpha,\alpha^{\dagger}\right\}  =-2\kappa\,.
\end{equation}
The quantum Noether charges generating translations and phase rotation
of the complex Grassmann plane parametrized by $\zeta$ are the  differential operators
\begin{equation}
\Pi = \partial_{\zeta}+\kappa\bar{\zeta}\, , \qquad
\Pi^\dagger =  \partial_{\bar{\zeta}}+\kappa\zeta\, ,\qquad
F= \zeta\partial_\zeta - \bar\zeta\partial_{\bar\zeta}\, .\label{GEN1}
\end{equation}
These span an `internal'  superalgebra for which the non-zero (anti)commutators are
\begin{equation}\label{fermsym}
\{\Pi, \Pi^\dagger\} = 2\kappa\, ,\qquad [F,\Pi ]= -\Pi\, ,
\qquad [F,\Pi^\dagger ]=\Pi^\dagger\,  .
\end{equation}
It is straightforward to check that these generators commute with
the Hamiltonian \p{fermham}. Note that the Hamiltonian can be written as
\be
H_f = \Pi^\dagger\Pi -2\kappa F -\kappa\,,\label{Sug0}
\ee
which implies that it belongs to the enveloping algebra of the superalgebra defined by the
relations (\ref{fermsym}).

A general wavefunction $\psi(\zeta,\bar\zeta)$ is a function of $\zeta$ and $\bar\zeta$,
which implies a total of four states. There are two ground states of energy $-\kappa$,
with wavefunction $\psi^{(0)}$ annihilated by $\alpha$, and two excited states of
energy $\kappa$, with wavefunction $\psi^{(1)}$ annihilated by $\alpha^\dagger$.
These energy eigenfunctions take the form
\begin{equation}
\psi^{(0)} = e^{-\kappa\zeta\bar\zeta}\, \psi_0\left(\zeta\right), \qquad
\psi^{(1)}= e^{\kappa\zeta\bar\zeta}\, \psi_1\left(\bar\zeta\right),\label{Psi1}
\end{equation}
for {\it analytic} function $\psi_0$ and {\it anti-analytic} function $\psi_1$:
\begin{equation}
\psi_0 =A_0 + \zeta\, B_0 \, ,\qquad \psi_1 = A_1 +  \bar\zeta\, B_1\,.
\label{vectors}
\end{equation}
The 2-vectors $(A_0\,, B_0)\,$, $(A_1\,, B_1)$ form two irreducible representations of
the supertranslation group defined above. Its generators have  the following
realization on $\psi_0$: $\Pi_0 = \partial_\zeta\,, \Pi^\dagger_0 = 2\kappa\zeta\,,
F_0 = \zeta\partial_\zeta\,$.

Now we must choose an inner product.  There are two obvious ways to proceed
and each is instructive. We consider them in turn.

\subsection{Superspace approach}

One natural choice of inner product  is
\begin{equation}
\left\langle \phi \big| \psi \right\rangle =\int d\zeta d\bar{\zeta}~\overline
{\phi\left(  \zeta,{\bar{\zeta}}\right)  }\psi\left(  \zeta,\bar{\zeta}
{}\right). \label{NaiveInnerProduct}
\end{equation}
This has the property that  $\alpha$ and $\alpha^{\dagger}$ are hermitian
conjugates, when viewed as operators on wavefunctions, which guarantees the hermiticity of
$H_f\,$. In turn, this guarantees the orthogonality of the energy eigenstates $\psi^{(0)}$ and
$\psi^{(1)}\,$. However, the product \p{NaiveInnerProduct} also implies a negative norm for excited states.
Indeed, one finds that
\begin{align}
\left\langle \psi^{(0)}\big|\psi^{(0)}\right\rangle  =2\kappa\bar{A}_{0}A_{0}
+\bar{B}_{0}B_{0}\,,\qquad
\left\langle \psi^{(1)}\big|\psi^{(1)}\right\rangle  =-2\kappa\bar{A}_{1}A_{1}
-\bar{B}_{1}B_{1}\,.
\end{align}
Therefore, with respect to the  inner product (\ref{NaiveInnerProduct})
for which the operators $\alpha$ and $\alpha^{\dagger}$ are conjugate to each
other and $H_{f}$ is manifestly hermitian, the norm is not positive
definite\footnote{Note that this is true irrespective of whether the $A$ and $B$ coefficients are
both ordinary complex numbers or complex supernumbers with  Grassmann-odd
products $AB$.}.

We can circumvent this difficulty by redefining the dual state vectors. Let us choose the
`metric' operator to be
\begin{equation}
G= - \kappa^{-1} H_f\,.
\end{equation}
Note that $G=G^\dagger$, as required, and that
\begin{equation}
G\left(\psi^{(0)}+\psi^{(1)}\right) = \psi^{(0)}-\psi^{(1)}\, ,
\end{equation}
which implies that the improved inner product is positive definite. As $G$ commutes
with the Hamiltonian in this example,
there are \emph{no} shifts introduced for the hermitian
conjugates of any of the symmetry generators, and hence no change in the
(anti)commutation relations (\ref{fermsym}).
However, the hermitian conjugation properties of non-conserved operators
can change. In particular, we have
\begin{equation}
\alpha^{\ddagger}=-\alpha^{\dagger}\, .
\end{equation}
The operators $\alpha$ and $\alpha^\ddagger$ have the commutation relations
of fermionic annihilation and creation operators,
\begin{equation}
\left\{ \alpha,\alpha^{\ddagger}\right\}  = 2\kappa\,,
\end{equation}
and the Hamiltonian is formally the same\footnote{The difference is the doublet
degeneracy of the two energy eigenstates, which is the fermionic version
of the infinite degeneracy of the energy levels of
the bosonic Landau model. This doublet degeneracy is related to the symmetry
under supertranslations, with algebra \p{GEN1}, just as the degeneracy in the bosonic Landau model
is related to the invariance under the `magnetic translations' defined in \p{PP}.}
as the hamiltonian of a fermionic harmonic oscillator:
\begin{equation}
H_{f}=\alpha^{\ddagger}\alpha-\kappa\,.
\end{equation}

Note that the conjugation properties of the coordinates and momenta \emph{are}
altered:
\begin{equation}
\zeta^{\ddagger}={\frac{1}{\kappa}}\partial_{\zeta}\ ,\qquad  \left(  \bar
{\zeta}\right)  ^{\ddagger}={\frac{1}{\kappa}}\partial_{\bar{\zeta}}\ ,
\end{equation}
and correspondingly
\begin{equation}
\left(\partial_\zeta\right)^\ddagger = \kappa\zeta\, ,\qquad
\left(\partial_{\bar\zeta}\right)^\ddagger = \kappa\bar\zeta\, .
\end{equation}
That is, under the new conjugation the momentum canonically conjugate
to a coordinate is also the coordinate's hermitian conjugate!

\subsection{Matrix approach}

A general wave function can be written as
\begin{equation}
\psi\left(  \zeta,{\bar{\zeta}}\right)  =\mathcal{A}+\zeta\mathcal{B}
+{\bar{\zeta}}\mathcal{C}+{\bar{\zeta}}\zeta\mathcal{D\ },
\end{equation}
for  constant complex coefficients $\mathcal{A},\mathcal{B},\mathcal{C}$ and $\mathcal{D}$.
In principle, these constants could be general super-numbers but we again suppose
{\it either} that they are ordinary complex numbers, in which case
the Hilbert space is $\bC^4$, {\it or} that $\psi$ is a superfield
(i.e. has definite Grassmann parity) in which case the `Hilbert' space
is  the supervector space $\bC^{(2|2)}$.  In either case, the action of $H_{f}$
is given by
\begin{equation}
H_{f}~\psi\left(  \zeta,{\bar{\zeta}}\right)  =-\mathcal{D}-\kappa
\zeta\mathcal{B}+\kappa{\bar{\zeta}}\mathcal{C}-\kappa^{2}{\bar{\zeta}}
\zeta\mathcal{A}\, .
\label{action}
\end{equation}

Clearly, any procedure involving this model  can be stated directly in terms
of $4\times4$ (super)matrices.  Let us associate to the superfield wavefunction
$\psi$ the column (super)vector\footnote{Note the non-alphabetic ordering.}
\begin{equation}\label{ColumnVector}
\vec\psi=\left(
\begin{array}
[c]{c}
\mathcal{A}\\
\mathcal{D}\\
\mathcal{B}\\
\mathcal{C}
\end{array}
\right).
\end{equation}
Independently of the grading assigned to the coefficients
$\mathcal{A}, \mathcal{B}, \mathcal{C},  \mathcal{D}$, the
differential operator $H_f$ is then equivalent to the (super)matrix
\begin{equation}
\mathcal{H}=\left(
\begin{array}
[c]{cccc}
& -1 &  & \\
-\kappa^2 &  &  & \\
&  &- \kappa & \\
&  &  & \kappa
\end{array}
\right).
\end{equation}
Because of its block-diagonal form, it is manifest that this may be viewed either
as a matrix or as a supermatrix.  In either case it is non-hermitian with respect
to the usual positive definite inner product
(for which $\mathcal{H}^{\dagger}=\overline{\mathcal{H}^{\text{T}}}$), but it is hermitian
with respect to the metric
\begin{equation}
\mathcal{G}=\left(
\begin{array}
[c]{cccc}
\kappa^2 &  &  & \\
& 1 &  & \\
&  & 1 & \\
&  &  & 1
\end{array}
\right),
\end{equation}
in the sense that $\mathcal{G}\mathcal{H}= \mathcal{H}^\dagger \mathcal{G}$; i.e. it is
`quasi-hermitian'  \cite{Scholtz1992}.

For any $\kappa\neq0$, the matrix $\mathcal{H}$ can be diagonalized
by a non-unitary similarity transformation,
$\mathcal{H=S}^{-1}\mathcal{H}_{\text{D}}\mathcal{S}$, and  the construction of a
positive definite inner product  in terms of the usual orthonormal basis
of the transformed system is then straightforward. The inverse similarity
transformation then leads from this orthonormal basis back to a bi-orthogonal
system (see the classic text \cite{Goursat}) which corresponds to the previous
polynomial basis and an appropriate set of dual polynomials. That is to say,
in terms of the original basis underlying (\ref{ColumnVector}), a suitable
metric is $\mathcal{G}=\mathcal{S}^{\dagger}\mathcal{S}$, where for example
\begin{equation}
\mathcal{S}=\left(
\begin{array}
[c]{cccc}
\kappa/\sqrt{2} &1/\sqrt{2} & & \\
- \kappa/\sqrt{2} & 1/\sqrt{2} &  & \\
& & 1& \\
 &  & & 1
\end{array}
\right).
\end{equation}
The matrix approach is thus equivalent to the superfield approach. However,  it is convenient only 
for finite-dimensional matrices, so we revert to the superspace approach in what follows. 

\section{The superplane model}
\label{sec:superplane}
\setcounter{equation}{0}

The Lagrangian of the superplane model \cite{Hasebe:2005cm,Ivanov:2005vh} is  the sum
$L= L_f+L_b$ of the Lagrangian \p{FermLL} of the fermionic Landau model  and the Lagrangian
\be
L_b = |\dot z|^2 - i\kappa\left(\dot z\bar z - \dot{\bar z} z\right),
\ee
of  Landau's original `bosonic' model,  where $2\kappa$ can now be identified
as the (positive) value of a uniform  magnetic field. The phase space Lagrangian is
\be\label{hamiltonianlag}
\tilde L = \left(\dot z p -i \dot\zeta\pi\right) + c.c.\  - \ H_{class}\,,
\ee
where
\be
H_{class} = \left|p+i\kappa\bar z\right|^2 +
\left(\bar\pi -\kappa\zeta\right)\left(\pi -\kappa \bar\zeta\right).
\ee
For a standard operator ordering prescription, the corresponding quantum Hamiltonian operator is
\be
H= \partial_{\bar\zeta}\partial_{\zeta}- \partial_{z}\partial_{\bar z} +
\kappa\left(  \bar z \partial_{\bar z} + \bar\zeta\partial_{\bar\zeta} - z
\partial_{z} - \zeta\partial_{\zeta}\right)  + \kappa^{2}\left(  z\bar z +
\zeta\bar\zeta\right).
\ee
Introducing the boson creation and annihilation operators
\begin{equation}\label{acbos}
a=i\left( \partial_{\bar{z}}+\kappa z\right)  \,,\qquad a^{\dagger}=i\left(
\partial_{z}-\kappa\bar{z}\right), \qquad \left[ a,a^\dagger\right] = 2\kappa\, ,
\end{equation}
and recalling the definition (\ref{fermiosc}) of the fermion creation and annihilation operators,
we find that
\be
H = a^{\dagger}a - \alpha^\dagger \alpha\, . \label{HamSP}
\ee
Note  the cancellation of the zero point energies.

The ground state  wavefunction $\psi^{(0)}$ for the lowest Landau level is annihilated
by both $a$ and $\alpha$ and hence takes the form
\be
\psi^{(0)} = e^{-\kappa K_2}\, \psi^{(0)}_{an}(z,\zeta)\,,  \ee
for {\it analytic} function $\psi^{(0)}_{an}\,$. Here we introduced the notation (see
\cite{Ivanov:2005vh})
\be\label{ktwo}
 K_2 = |z|^2 + \zeta{\bar \zeta}\, .
\ee
{}For each ground state, there are two excited states at the first Landau level, with
wavefunctions given by the action of {\it either} $a^\dagger$ {\it or} $\alpha^\ddagger$
on the ground-state wavefunction.
The wavefunctions at higher Landau levels (with the energy $E_N = 2\kappa N$) are obtained
similarly and have the same degeneracy.
Thus the $N$th level
Hilbert space has a wavefunction of the form
\be\label{NSP}
\psi^{(N)} = \left(-ia^\dagger\right)^N e^{-\kappa K_2} \psi^{(N)}_+ \left(z,\zeta\right)
- N \left(-ia^\dagger\right)^{N-1}\alpha^\dagger e^{-\kappa K_2}\psi^{(N)}_-\left(z,\zeta\right),
\ee
where $\psi_\pm(z,\zeta)$ are two analytic functions of $z$ and $\zeta$, and the factors of $i$ and $N$
are included for convenience of comparison with our later results. We may write
these analytic wavefunctions as
\be \label{decPM}
\psi^{(N)}_\pm(z,\zeta) = A^{(N)}_\pm (z) + \zeta B^{(N)}_\pm (z)\,,
\ee
where the $A$ and $B$ coefficients are now analytic functions of $z$; a four-fold degeneracy
of the excited states, relative to the bosonic Landau model, is now manifest.

The Hamiltonian commutes with the `magnetic translation' operators
\begin{equation}
P=-i\left(  \partial_{z}+\kappa\bar{z}\right)  \,,\qquad P^{\dagger}=-i\left(
\partial_{\bar{z}}-\kappa z\right)  \label{PP}
\end{equation}
and with the supermagnetic translation operators $(\Pi,\Pi^\dagger$) defined in \p{GEN1}. The non-zero
(anti)commu- tation relations of these supertranslation operators are
\begin{equation}
\lbrack P,P^{\dagger}]=2\kappa\,,\qquad \{\Pi^{\dagger},\Pi\}=2\kappa\,.
\label{100}
\end{equation}
The Hamiltonian also commutes with the operators:
\begin{equation}
Q=z\partial_{\zeta}-\bar{\zeta}\partial_{\bar{z}}\,,\qquad  Q^{\dagger}=\bar
{z}\partial_{\bar{\zeta}}+\zeta\partial_{z}\, , \label{QQbar}
\end{equation}
and
\begin{equation}
C=z\partial_{z}+\zeta\partial_{\zeta}-\bar{z}\partial_{\bar{z}}-\bar{\zeta
}\partial_{\bar{\zeta}}\, .
 \label{Cdef}
\end{equation}
These operators span the algebra of $SU(1|1)$, for which
the only non-zero (anti)commu- tation relation is
\begin{equation}
\{Q,Q^{\dagger}\}=C\,. \label{101}
\end{equation}
Including the operators $P,\ P^{\dagger},\Pi$ and $\Pi^{\dagger}$ leads to the
semi-direct product superalgebra $ISU(1|1)\,$.  In particular,
\begin{equation}
[Q,P]=i\Pi\, ,\qquad \{Q^\dagger,\Pi\}=iP\,,\qquad [C,P]
=-P\,,\qquad  [C,\Pi]=-\Pi \,. \label{102}
\end{equation}
For the Hamiltonian \p{HamSP} there exists a representation in terms of
the $ISU(1|1)$ charges, analogous to \p{Sug0}:
\begin{equation}
H = P^{\ddagger}P + \Pi^{\ddagger}\Pi- 2\kappa C\,. \label{Sug2}
\end{equation}

\subsection{Norm and  modified $ISU(1|1)$ algebra}
\label{subsec:norm}

The natural  $ISU(1|1)$-invariant inner product is such that states at different levels
are orthogonal and states within the same level have inner product
\begin{equation}\label{innersp}
\left\langle \phi \big|\psi\right\rangle =\int \!d\mu\
\overline{\phi\left( z,\bar z;\zeta, \bar \zeta\right)}\,
\psi\left(z,\bar z; \zeta,\bar\zeta\right),
\end{equation}
where $d\mu$ is the  $ISU(1|1)$-invariant superspace measure
\be\label{measure}
d\mu = dzd\bar{z}d\zeta d\bar{\zeta}\, .
\ee
As in the purely fermionic case, and for the same reason, this leads to negative norm states.
Specifically, one finds that
 \be\label{indefnorm}
\langle\psi^{(N)}\big| \psi^{(N)} \rangle \  = \ (2\kappa)^N N! \left[
-N \left|\left| \psi_- ^{(N)}\right|\right|^2 +   \left|\left| \psi_+^{(N)}\right|\right|^2\right],
\ee
where we have defined
\be\label{psi}
\left|\left| \phi_{an}\right|\right|^2  \equiv
\int \!d\mu \ e^{-2\kappa K_2} \ \overline{\phi_{an}} \ \phi_{an}
\ee
for any {\it analytic} function, or superfield,  $\phi_{an}(z,\zeta)$. A computation
shows that\footnote{Recall that either the $A$ or the $B$ coefficient function will be
Grassmann odd if the wavefunction is a superfield.}
\be
\left|\left| \psi_\pm^{(N)} \right|\right|^2 = \int \!dzd\bar z\,
e^{-2\kappa|z|^2}\left( 2\kappa \overline{A^{(N)}_{\pm}(z)}
A^{(N)}_{\pm}(z) + \overline{B^{(N)}_{\pm}(z)}
B^{(N)}_{\pm}(z)\right),\label{psi2}
\ee
so the minus sign in (\ref{indefnorm}) implies an indefinite norm. This
problem is circumvented exactly as before, and with the same metric operator
$G= - \kappa^{-1}H_f$, which we may write as
\begin{equation}\label{Gone}
G=\frac{1}{\kappa}\left[  \partial_{\zeta}\partial_{\bar\zeta} +\kappa^{2}
\bar\zeta\zeta+ \kappa\left(  \zeta\partial_{\zeta}- \bar\zeta\partial
_{\bar\zeta}\right)  \right].
\end{equation}
It is evident that $G$ commutes with $H$. It is also easy  to verify that $G$
commutes with the operators $a$ and $a^{\dagger}$, but \emph{not} with
$\alpha$ and $\alpha^{\dagger}$, and this leads to the modified hermitian conjugates
\begin{equation}
\alpha^{\ddagger}=-\alpha^{\dagger}\, , \label{Qddagger}
\end{equation}
as in the fermionic Landau model. The Hamiltonian may now be written in
the manifestly positive form
\begin{equation}\label{Gsp}
H= a^\dagger a + \alpha^\ddagger \alpha\, .
\end{equation}

The metric operator $G$ commutes with all the bosonic symmetry generators of $ISU(1|1)$,
and the fermionic generators $\Pi$ and $\Pi^{\dagger}$, which therefore
all have unchanged hermitian conjugates. However $G$ does \emph{not} commute with $Q$,
and this leads to the modified hermitian conjugate
\begin{equation}\label{newQdag}
Q^{\ddagger}=Q^{\dagger}- \frac{i}{\kappa}\,  S\,,
\end{equation}
where the \emph{shift} operator is
\be
S=  i\left(  \partial_{z}\partial_{\bar\zeta} +
\kappa^{2} \bar z \zeta-\kappa\bar z \partial_{\bar\zeta} - \kappa
\zeta\partial_{z}\right).
\ee
As explained in section 2, it is convenient to introduce the new operator
\begin{equation}
\tilde Q=Q -{\frac{i}{2\kappa}}S^{\ddagger} \, , \label{Q1Q}
\end{equation}
since this operator commutes with $G$ and therefore has the property that
$\tilde Q^\ddagger = \tilde Q^\dagger$. We now have
\begin{equation}\label{centralH}
\left\{ \tilde Q,\tilde Q^{\dagger}\right\}  = \tilde C\,,
\end{equation}
where
\be\label{ctilde}
\tilde C= C+{\frac{1}{2\kappa}}H\, .
\ee
We now have two commuting symmetries, one an $ISU(1|1)$ symmetry with the modified charges
$(P, \Pi,\tilde Q; P^\dagger ,\Pi^\dagger,\tilde Q^\dagger; \tilde C)$,
and the other a worldline supersymmetry algebra with charges $(S,S^\ddagger;H)$. The generator
$\tilde{C}$ differs from the original $C$ by the term proportional to $H$, which commutes with all
symmetry generators and so can be thought of as a central charge. Thus the new $ISU(1|1)$
algebra can be interpreted as a central extension of the original $ISU(1|1)$ algebra.

\subsection{Worldline supersymmetry}

As explained in Section \ref{sec:prelim}, the hermiticity of the Hamiltonian
with respect to both the original and the modified norm implies that both $S$ and
$S^\ddagger$ are  constants  of motion. These operators
can be written as
\begin{equation} \label{susy}
S= a^{\dagger}\alpha\, ,\qquad S^\ddagger = a\alpha^\ddagger\, ,
\end{equation}
and they  have the anticommutation relation
\begin{equation}\label{super}
\{S,S^\ddagger \} = 2\kappa H\, ,\qquad \{S,S\}=0=\{S^\ddagger,S^\ddagger\}\, ,
\end{equation}
which is an  ${\cal N}{=}2$ worldline supersymmetry algebra. Note also that
\be
\{S,\tilde Q\}=0 \, ,\qquad \{S,\tilde Q^\dagger\}=0\, .
\ee
The worldline supersymmetry is unbroken because the ground state
is annihilated by both $S$ and $S^\ddagger$. The ground state is a singlet of the
${\cal N}{=}2$ worldline supersymmetry, but still forms a non-trivial multiplet
of $ISU(1|1)$, which explains its doublet degeneracy.
All higher $N$ states form non-trivial multiplets of ${\cal N}{=}2$ worldline supersymmetry
consisting of two irreducible $ISU(1|1)$ multiplets. This implies the four-fold
degeneracy of these states\footnote{Of course, these degeneracies should be understood
as relative to the bosonic Landau model.}.

Classically, the supersymmetry charges generate transformations of the phase-space variables.
After elimination of the momentum variables one finds that the infinitesimal transformation
generated by $\epsilon S + \bar\epsilon S^\ddagger$, for complex anticommuting
parameter $\epsilon$, is
\be
\delta z = \epsilon \dot\zeta  \, ,\qquad \delta \zeta = - \dot z \bar\epsilon\, .\label{SUSYtran}
\ee
It is readily verified that the configuration space Lagrangian is invariant under
these transformations, and that their algebra closes, on-shell, to the worldline supersymmetry
algebra. This classical supersymmetry is unbroken by the classical ground state solutions,
for which both $z$ and $\zeta$ are constant, as expected from the fact that worldline
supersymmetry is unbroken quantum mechanically.
The worldline supersymmetry is quite remarkable, taking into account the unconventional
form of the above transformations; conventionally, $z$ would vary into some fermionic field of the
proper dimension, not into its time derivative.
This unconventional form means that the commutator on $z$ and $\zeta$ involves $\ddot z$ and $\ddot \zeta$
rather than $\dot z$ and $\dot \zeta$; but $\ddot z$ and $\dot z$ are related by
the equations of motion, as are $\ddot \zeta$ and $\dot \zeta$, this being
a characteristic feature of Landau
models. For this reason, the on-shell closure of \p{SUSYtran} involves $\kappa$, so the
${\cal N}{=}2$ supersymmetry is made possible by the WZ terms with non-zero coefficient
$\kappa$.

Although worldline supersymmetry has emerged as an `accidental' symmetry in the
sense that it played no role in the construction of the model, there is another sense
in which it is `almost'  built into the construction.  This follows from the observation that
the worldline supersymmetry generators belong to the enveloping algebra of
$ISU(1|1)$, as is shown by the Sugawara-type representation
\begin{equation}
S=2i\kappa Q^{\ddagger}+P\Pi^{\ddagger}\ ,\qquad S^{\ddagger}
=-2i\kappa Q+P^{\ddagger}\Pi  \label{Sugawara1}
\end{equation}
and \p{Sug2}. The anticommutation relations of (\ref{super}) are now a direct consequence
of the $ISU(1|1)$ (anti)commutation relations of (\ref{100}), (\ref{101}), (\ref{102}).


It may appear from this result that worldline supersymmetry is an automatic consequence
of $ISU(1|1)$ symmetry, but this is not quite true. Suppose that we try to similarly
define supercharges $\tilde S$ and $\tilde S^\ddagger$ in terms of the modified
$ISU(1|1)$ generators. We then have
\be
\tilde S = 2i\kappa \tilde Q^\ddagger + P\Pi^\ddagger\, ,\qquad
\tilde S^\ddagger = -2i\kappa \tilde Q + P^\ddagger \Pi
\ee
and
\be
\tilde H= P^\ddagger P + \Pi^\ddagger \Pi -2\kappa \tilde C\, .
\ee
However, {\it these charges are identically zero}, as a consequence of the further
Sugawara-type relations\footnote{These relations show that the modified $ISU(1|1)$ supersymmetry
belongs to the enveloping algebra of the superplane translation algebra.}
\be
\tilde Q = -\frac{i}{2\kappa}\, P^\dagger \Pi\, ,\qquad
\tilde Q^\dagger = \frac{i}{2\kappa}\, P \Pi^\dagger\, ,
\ee
and
\be
\tilde C= \frac{1}{2\kappa} \left[P^\ddagger P + \Pi^\ddagger \Pi \right].
\ee

Thus, worldline supersymmetry is not an {\it automatic} consequence of $ISU(1|1)$ invariance,
and this was the reason for the qualification `almost'. In fact,  it should be obvious
that the Sugawara construction cannot yield anything new in a `natural' basis for the charges
which makes manifest that the symmetry group is the direct product of $ISU(1|1)$
and worldline supersymmetry. So the apparently miraculous construction of the worldline
supersymmetry algebra from the $ISU(1|1)$ algebra is really just a consequence of the
fact that we did not initially obtain the generators in their natural basis.

In order to better understand the origin of worldline supersymmetry, we now turn
to the planar superflag models.

\section{The planar superflag model}
\setcounter{equation}{0}

The superflag Landau model \cite{Ivanov:2004yw} describes a charged particle on the coset
superspace $SU(2|1)/[U(1)\times U(1)]$. One of the two Wess-Zumino (WZ) terms
associated with the $U(1)\times U(1)$ group is the Lorentz coupling to
a uniform magnetic field of strength $2\kappa\,$, where $\kappa$ can be identified
as the constant already introduced in the previous sections. The second WZ term
is a purely `fermionic'  one with constant coefficient $M$. The details may be
found in  \cite{Ivanov:2004yw};  here we are concerned with the planar limit,
in which one finds the following $ISU(1|1)$-invariant Lagrangian \cite{Ivanov:2005vh}:
\bea\label{planarSF}
L &=& \left(1+\bar\xi\xi\right)|\dot z|^2 + \left(\bar\xi \dot{\bar z} \dot \zeta
- \xi \dot z \dot{\bar\zeta}\right) + \bar\xi\xi \dot\zeta \dot{\bar\zeta}  \nonumber\\
&&- \ i\kappa\left(\dot z \bar z - \dot{\bar z} z+ \dot\zeta\bar\zeta
+ \dot{\bar\zeta} \zeta\right)
+iM\left(\bar\xi\dot\xi + \xi\dot{\bar\xi}\right).
\eea
Notice that this becomes the superplane Landau model Lagrangian of (\ref{hamiltonianlag})
when all terms involving the new anticommuting variable $\xi$ are omitted.  Notice too
that $\xi$ is auxiliary when $M{=}0$;  its elimination returns us to the superplane
Lagrangian\footnote{Assuming that $\dot z\ne0$; this is a subtlety dealt with
in \cite{Ivanov:2005vh}, where the quantum equivalence of the $M{=}0$
planar superflag model to the superplane model was established.} so we now have
a one-parameter deformation of the superplane Landau model  that both preserves
the $ISU(1|1)$ symmetry and retains the property  that the bosonic truncation yields
Landau's original model. The new variable $\xi$ was interpreted in  \cite{Ivanov:2005vh}
as a Nambu-Goldstone variable associated with the spontaneous breaking of
the $ISU(1|1)$ `supersymmetry', generated by the Noether charge $Q$. However, we
shall see (at least for $M{<}0$) that its interpretation in the quantum theory
with positive norm is as a Nambu-Goldstone variable for the spontaneous breakdown of
an  ${\cal N}{=}2$ worldline supersymmetry.

It will be instructive  to consider the classical theory before turning to the quantum theory.
Introducing the momentum variables  $(p, \pi)$  conjugate to $(z,\zeta)$, we
can express the Lagrangian in the alternative form\footnote{This is essentially eq. (3.7) of
\cite{Ivanov:2005vh} after using the phase space constraint $\varphi_\xi\approx 0$ to
eliminate the momentum variable $\chi$ conjugate to $\xi$,  but with $\tilde p$ of that
reference written here as $p$.}
\be\label{planarSFalt}
L = \left\{\left[\dot z p -i \dot\zeta\pi -iM \dot\xi\bar\xi \right] + \lambda\varphi \right\}
+c.c \ - H_{class}\,,
\ee
where
\be
H_{class} = \left(1-\bar\xi\xi\right)\left|p+i\kappa \bar z\right|^2
\ee
is the classical Hamiltonian, and $\lambda$ is a Lagrange multiplier for
the constraint $\varphi\approx 0$, where
\be
\varphi = \pi -\kappa\bar\zeta +i\bar\xi\left(p+i\kappa\bar z\right).
\ee
If this constraint is used to eliminate $\pi$, we get a Lagrangian in terms
of the complex variables $(z,\zeta,\xi,p)$, for which the Euler-Lagrange equations
are equivalent to
\bea
\dot z  &=& \left(1+\frac{i}{2\kappa} \bar\xi\dot\xi\right)\left(\bar p -i\kappa z\right), \qquad
\dot p = i\kappa \dot{\bar z}\, , \nonumber\\
\dot\zeta &=& -\left[\xi + \frac{i}{2\kappa}\dot\xi
\left(1-\bar\xi\xi\right)\right]\left(\bar p -i\kappa z\right),
\eea
and
\be\label{xieom}
\left[H_{class}-4\kappa M\right]\dot\xi=0\, .
\ee
This last equation shows that $\dot\xi=0\,$, except when the energy equals $4\kappa M$.
This is never the case when $M{<}0$, so the equations of motion for $M{<}0$ are equivalent to
\be
\dot z = \left(\bar p -i\kappa z\right)\, , \qquad \dot p =i\kappa\dot{\bar z}\, ,\qquad
\ee
and
\be
\dot\zeta = -\xi \dot z\, , \qquad \dot\xi=0\, .
\ee
These equations imply the superplane Landau model equations of motion
\be
\ddot z = -2i\kappa \dot z\, , \qquad \ddot\zeta =-2i\kappa \dot \zeta\, .
\ee

However, whereas the initial conditions for the superplane model are the values at
a given time of $(z,\dot z, \zeta,\dot\zeta)\,$, the initial conditions for the equations
of the $M{<}0$ planar superflag model are the values  of $(z,\dot z, \zeta,\xi)\,$. These are
equivalent as long as $\dot z\neq 0\,$, because then $\xi=-\dot \zeta/\dot z\,$
but they are inequivalent at $\dot z=0\,$. Specifically, $\dot z=0$ implies $\dot\zeta=0$
for the planar superflag model, but $\xi$ is then undetermined. This implies that $\xi$
is an independent, albeit  constant, variable in a classical ground state, for which $H_{class}=0\,$.
This is also true for $M=0$ (where $\xi$ is auxiliary for  $\dot z\neq 0$) but in this case
(i) $\xi$ need not be constant  because (\ref{xieom}) is an identity when $H_{class} \sim |\dot z|^2 =0\,$, and (ii) $\xi(t)$ can be `gauged away' by  a fermionic gauge invariance, as shown in  \cite{Ivanov:2005vh} (where it was also shown that a similar gauge invariance arises when $2M$ is any non-negative integer). The significance of these facts will become apparent when
we discuss worldline supersymmetry, but let us stress here the independence of
the classical physics on $M$ as long as $M{<}0\,$. As we should expect, we will find that
the same is true of the quantum theory.

\subsection{Quantum theory}

The quantization of the planar superflag model is complicated by the phase-space constraint.
In particular, the classical Hamiltonian $H_{class}$ does not have weakly vanishing Poisson brackets
with the constraint function $\varphi$ and its complex conjugate $\bar\varphi$.
This problem was dealt with in  \cite{Ivanov:2005vh}  by a  change of variables but it
was noted that an alternative approach would be to consider the modified
Hamiltonian\footnote{This is eq. (3.22) of \cite{Ivanov:2005vh}  but with $\tilde p$ now written as $p$.}
\be
H'_{class}= \left(1+ \bar\xi\xi\right) \left| p + i\kappa \bar z
+ i\xi \left(\pi - \kappa\bar\zeta\right)\right|^2,
\ee
which is weakly equal to $H_{class}$  and has weakly vanishing Poisson brackets with
both $\varphi$ and $\bar\varphi$. This alternative approach is much more convenient
for  present  purposes. The results obtained in this way are of course equivalent
to those of  \cite{Ivanov:2005vh}, but the wavefunctions  are now functions of $(z,\zeta,\xi)$.
Following \cite{Ivanov:2005vh}, we define
\be
K_1 = 1+ \bar\xi\xi
\ee
and introduce the `shifted'  $z$ variable
\be
z_{sh} = z+ \bar\xi \zeta \, , \qquad  \bar z_{sh} = \bar z - \xi\bar\zeta\, .
\ee
We may now quantize without constraint  provided that we restrict to
`physical'  wavefunctions, which take the form
\be
\Psi = K_1^M e^{-\kappa K_2} \Psi_{ch}\left(z,\bar z_{sh}, \zeta,\xi\right),
\ee
where $\Psi_{ch}$ is a `chiral'  wavefunction that depends
on $\bar\zeta$ only through $\bar z_{sh}$, and $K_2$ was defined in (\ref{ktwo});
we refer to \cite{Ivanov:2005vh} for details. The Hamiltonian operator acting
on these wavefunctions can be written as
\be\label{qhamsf}
H= \hat a^\dagger \hat a\,,
\ee
where the  `non-linear' annihilation and creation operators
\be\label{ahats}
\hat a = i\sqrt{K_1}\left(\partial_{\bar z} + \kappa\, z_{sh}
- \bar\xi\partial_{\bar\zeta}\right),\qquad
\hat a^\dagger = i\sqrt{K_1}\left(\partial_z
- \kappa\, \bar z_{sh}- \xi\partial_\zeta \right),
\ee
have the same commutation relation as for the bosonic Landau model:
\be
\left[ \hat a,\hat a^\dagger\right] = 2\kappa \,.
\ee
In writing the Hamiltonian operator as (\ref{qhamsf}) we are
resolving the operator ordering ambiguity by a `normal ordering' prescription
that differs from the `harmonic oscillator' prescription that we used for the
bosonic Landau model (and in \cite{Ivanov:2005vh}). As a consequence,
$H$ has eigenvalues $2\kappa N\,$, where $N$ is a non-negative integer,
exactly as for the superplane Landau model.
In the physical energy eigenfunctions at level $N$ the chiral
wavefunction is expressed through an {\it analytic} function
of $(z,\zeta,\xi)$ as
\be
\Psi_{ch}^{(N)} = \tilde\nabla_z^N \Psi_{an}^{(N)}\left(z,\zeta,\xi\right), \qquad
\tilde\nabla_z = \partial_z -2\kappa \bar z_{sh} - \xi\partial_\zeta\,.
\ee

It is useful to note that any physical operator  ${\cal O}$ is defined by
its action on the energy eigenfunctions $\Psi^{(N)}$, and if it commutes
with the Hamiltonian then this action is determined by an associated
`short' operator ${\cal O}_{an}$ acting on the associated analytic
wavefunctions $\Psi_{an}^{(N)}$:
\be\label{fullshort}
{\cal O}\, \Psi^{(N)} =  K_1^M \,e^{-\kappa K_2}\tilde\nabla_z^N \,
{\cal O}_{an} \Psi^{(N)}_{an}\, .
\ee
In particular, the short form of the Hamiltonian operator is
\be
H_{an}= 2\kappa N_{an}\, ,
\ee
where $N_{an}$ is the `short'  level number operator defined by
\be
N_{an}\Psi_{an}^{(N)} = N\, \Psi_{an}^{(N)}\, .
\ee
As the operators generating the $ISU(1|1)$ symmetry commute with
the Hamiltonian, they too may be represented by their short forms, which are
\begin{align}\label{pshorter}
P_{an} &= -i\partial_z \, ,\qquad P_{an}^\dagger = 2i\kappa z\,, \nonumber \\
\Pi_{an} &= \partial_\zeta\, ,\qquad \Pi_{an}^\dagger = 2\kappa\zeta\, ,\nonumber \\
Q_{an} &= z\partial_\zeta - \partial_\xi \, , \qquad
Q_{an}^\dagger = \zeta\partial_z + \left(N_{an}-2M\right)\xi \, ,\nonumber\\
C_{an} &= \zeta\partial_\zeta + z\partial_z+ 2M -N_{an} \, .
\end{align}
One may verify that the associated operators
$(P,P^\dagger;\Pi,\Pi^\dagger; Q, Q^\dagger; C)$,  defined via (\ref{fullshort}),
satisfy the $ISU(1|1)$ (anti)commutation relations (\ref{100}), (\ref{101}) and (\ref{102}), and that 
$(P^\dagger, \Pi^\dagger, Q^\dagger)$ are the hermitian conjugates of $(P,\Pi,Q)$,
with respect to the $ISU(1|1)$-invariant  inner product:
\be\label{innerch}
\langle \Phi \big| \Psi\rangle = \int \!d\mu \int\! d\xi d\bar\xi\ \overline{\Phi}\, \Psi \
= \int \! d\mu\,  e^{-2\kappa K_2}\! \int \! d\xi d\bar\xi\ K_1^{2M}\,
\overline{\Phi_{ch}} \, \Psi_{ch}\, ,
\ee
where $d\mu = dz d\bar z d\zeta d\bar\zeta$ is the measure of (\ref{measure}).
More generally, for any `physical' operator ${\cal O}$ (i.e. one that acts on
`physical' wavefunctions), we define ${\cal O}^\dagger$ to  be its hermitian
conjugate with respect to this inner product.

When acting on physical states, the Hamiltonian can be written as
\begin{equation}\label{sugH}
H = P^{\ddagger}P + \Pi^{\ddagger}\Pi- 2\kappa C + 4\kappa M\, , 
\end{equation}
which is analogous to (\ref{Sug2}), with which it coincides for $M=0$. It follows that H
is hermitian with respect to the above inner product, and the hermiticity of H implies that  energy eigenfunctions in different Landau levels are orthogonal. The `superflag' norm of an energy
eigenfunction within the $N$th level  is given by
\be\label{sfnorm}
\left|\left| \Psi^{(N)}\right|\right|^2_{sf} \equiv \langle\Psi^{(N)} \big| \Psi^{(N)}\rangle =
 \int \! d\mu\,  e^{-2\kappa K_2}\! \int \! d\xi d\bar\xi\ K_1^{2M}\,
 \left| \tilde\nabla_z ^N \Psi_{an}^{(N)}\right|^2.
\ee
We may write
\be\label{PanN}
\Psi_{an}^{(N)} = \psi_-^{(N)} \left(z,\zeta\right) + \xi\,  \psi_+^{(N)}\left(z,\zeta\right)
\ee
and
\be
\tilde \nabla_z = \tilde D_z + \xi\left(2\kappa\bar\zeta - \partial_\zeta\right), \qquad
\tilde D_z = \partial_z -2\kappa \bar z\, ,
\ee
to get
\be
\tilde\nabla^N \Psi^{(N)}_{an} = \tilde D_z^N\psi_-^{(N)} + \xi \left[\tilde D_z^N\psi_+^{(N)} +
N \tilde D_z^{N-1}\left(2\kappa\bar\zeta - \partial_\zeta\right)\psi_-^{(N)}\right].
\ee
Performing the Berezin integration over $\xi$ and $\bar\xi$ in (\ref{sfnorm}) then gives
\bea
\left|\left| \Psi^{(N)}\right|\right|^2_{sf} &=&  \int \!d\mu \, e^{-2\kappa K_2} \bigg\{
2M\left| \tilde D_z^N \psi_-^{(N)}\right|^2 + \nonumber \\
&&+\ \left|\tilde D_z^{(N)}\psi_+^{(N)}
+ N  \left(2\kappa\bar\zeta -
\partial_\zeta\right)\tilde D_z^{N-1} \psi_-^{(N)}\right|^2\bigg\}\,.
\eea
The cross term in the expansion of the final term in this expression is zero,
as can be proved by integration by parts of the term with $\partial_\zeta\,$. We thus have
\bea
\left|\left| \Psi^{(N)}\right|\right|^2_{sf} &=&  \int \!d\mu \, e^{-2\kappa K_2} \bigg\{
2M\left| \tilde D_z^N \psi_-^{(N)}\right|^2 + \left|\tilde D_z^{(N)}\psi_+^{(N)}\right|^2
 \nonumber \\
&&+\ N^2 \left|\left(2\kappa\bar\zeta - \partial_\zeta\right)\tilde D_z^{N-1} \psi_-^{(N)}\right|^2
 \bigg\}\,.
\eea
One may further show by integration by parts that
\be
\int \!d\mu \,  e^{-2\kappa K_2} \left|\left(2\kappa\bar\zeta - \partial_\zeta\right)
\tilde D_z^{N-1} \psi_-^{(N)}\right|^2
= -2\kappa \int \!d\mu \, e^{-2\kappa K_2} \left|\tilde D_z^{N-1} \psi_-^{(N)}\right|^2
\ee
and also that
\be
 \int \!d\mu \, e^{-2\kappa K_2} \left| \tilde D_z^J \psi_\pm^{(N)}\right|^2
 = (2\kappa)^J J! \int \!d\mu \, e^{-2\kappa K_2} \left| \psi_\pm^{(N)}\right|^2,
 \ee
 for any integer $J$. We thus find that
 \be\label{5.23}
\left|\left| \Psi^{(N)}\right|\right|_{sf}^2 = (2\kappa)^N N!\left[
(2M-N)\left|\left| \psi_- ^{(N)}\right|\right|^2
+   \left|\left| \psi_+^{(N)}\right|\right|^2\right],
\ee
where the norm on the right hand side is the `analytic-function norm' defined in
(\ref{indefnorm}) and given explicitly for the $\psi_\pm$ analytic functions in (\ref{psi}),
\p{psi2}. This is the result of  \cite{Ivanov:2005vh}.  With this norm,  there are ghosts
 in the levels with $N{>}2M$, and zero-norm states in the level with $N{=}2M$
 whenever $2M$ is a non-negative integer. Note the agreement with (\ref{indefnorm})
 for $M=0$, which is a consequence of the equivalence of the $M=0$ model with the
 superplane model for the `naive' superspace norm.

\subsection{Positive inner product}

The inner product  (\ref{innerch}) is not unique but if we wish to
preserve the $ISU(1|1)$ invariance  then any planar superflag metric operator $G_{sf}$ yielding
a new inner product must be a function only of $\xi$ and $\partial_\xi$.
If we also require that $G_{sf}$ have even Grassmann parity and is such that $G_{sf}^2=1$,
then there are only two possibilities for its `short' form:  either $G_{an}=1$,
which implies $G_{sf}=1$ (as in \cite{Ivanov:2005vh} and assumed so far), or
\be\label{newG}
G_{an} = [\xi,\partial_\xi] = -1 + 2\xi\partial_\xi\, .
\ee
One may verify that the corresponding operator $G_{sf}$ has all the properties
required of a metric operator. Observing that
\be
G_{an} \Psi_{an}^{(N)} = - \psi_-^{(N)} + \xi\, \psi_+^{(N)}\, ,
\ee
we deduce that the new norm of $\Psi^{(N)}$ is
\be\label{invnorm2}
 \langle \langle\Psi^{(N)}\big|
\Psi^{(N)}\rangle\rangle   \equiv  \langle\Psi^{(N)}\big| G_{sf}\, \Psi^{(N)}\rangle
\propto \left(N-2M\right) \left|\left| \psi_-^{(N)}\right|\right|^2
+  \left|\left|\psi_+^{(N)}\right|\right|^2.
\ee
All states now have positive norm when $M{<}0$. This remains true for $M{=}0$ except
that half of the $N{=}0$ states, namely those comprised by $\Psi^{(0)}_-$,  have zero norm.
When there are zero-norm states, the vector (super)space of physical states is the
quotient of the space of all states by the subspace of zero-norm states,
which means that any state of zero-norm  corresponds to the zero-vector of the physical space.
Thus, zero-norm states do not contribute to the physical spectrum. Taking this into account,
it follows that the $M{=}0$ planar superflag model has precisely the same spectrum,
including degeneracies, as the superplane model, and is therefore equivalent to
it.

In view of this equivalence, our choice of superflag metric operator $G_{sf}$
should imply, for $M{=}0$,
the superplane metric operator $G$ of (\ref{Gone}). To verify this, we note that
the superflag wavefunction $\Psi^{(N)}$ has the $\xi$-expansion
\bea
\Psi^{(N)} &=& \left(-ia^\dagger\right)^N e^{-\kappa K_2}\psi_-^{(N)} \nonumber\\
&&+\  \xi\, \left[ \left(-ia^\dagger\right)^N e^{-\kappa K_2}\psi_+^{(N)}
- N \left(-ia^\dagger\right)^{N-1}\alpha^\dagger e^{-\kappa K_2} \psi_-^{(N)}\right],
\eea
where $a^\dagger$ and $\alpha^\dagger$ are the superplane creation operators introduced in
(\ref{fermiosc}) and (\ref{acbos}). Noting that
\be
\partial_\xi \Psi^{(N)} = \psi^{(N)}\, ,
\ee
where $\psi^{(N)}$ is precisely the superplane energy eigenfunction of (\ref{NSP}),  we see that
\be
\int \!d\mu  \int \!d\xi d\bar\xi \ \overline{\Psi^{(N)}}\,   \Psi^{(N)}
= \int d\mu\   \overline{\psi^{(N)}}\, \psi^{(N)}\, ,
\ee
and hence the `naive' $M=0$ superflag norm coincides with the `naive' superplane norm, as expected.
We now observe that
\be
\partial_\xi \left(G_{sf}\Psi^{(N)} \right) = G \, \psi^{(N)}\, ,
\ee
from which it follows that the modified planar superflag norm implies the modified norm introduced
earlier for the superplane model.

When $M{>}0$ there are negative-norm states for all $N{<}2M$, in particular for $N{=}0$,
but one can
revert to the `naive'  norm for these levels, thus ensuring that all states
have a positive-definite (or zero) norm for any value of $M$. Note that the two norms
coincide when $N{=}2M$, which can happen only when $2M$ is a non-negative integer,
and in this case there are zero-norm states. The $M{=}0$ case discussed above is just a special
case of this phenomenon. Taking into account the possibility of zero-norm states,
we see that the spectrum is the same for all $M$, with the same degeneracy
at each Landau level, except when $2M$ is a non-negative integer. Every
such non-negative integer yields a different spectrum because half of
the states in the $N{=}2M$ level have zero norm.  In what follows we assume
that $M{<}0$ so that the metric operator $G_{sf}$ is given by (\ref{newG});
the modification required for the $N{<}2M$ states when $M{>}0$ will be
obvious since the `naive' norm then applies.

The only $ISU(1|1)$ generators that fail to commute with $G_{sf}$ are $Q$ and $Q^\dagger$:
\be\label{Gcomm}
\left[  G_{an} ,Q_{an}\right]  =2\partial_\xi\, ,\qquad
\left[  G_{an},Q_{an}^{\dagger}\right]  =2\xi \left(N_{an} -2M\right).
\ee
This means that all hermitian conjugates are unmodified except for those of $Q$
and $Q^\dagger$.
Following the general procedure, we have
\begin{align}
Q^{\ddagger}_{an} &  = G_{an}\,Q^{\dagger}_{an}\,G_{an} = \left(
Q^{\dagger}_{an} - \frac{i}{\kappa} S_{an}\right),\\
(Q^{\dagger}_{an})^{\ddagger}&  = (Q^{\ddagger}_{an})^{\dagger
} =G_{an}\,Q_{sh}\,G = \left(  Q_{an} + \frac{i}{\kappa}
S^{\dagger}_{an}\right),
\end{align}
whence, using (\ref{Gcomm}), the shift operators $S$ and $S^{\dagger}$ are
found to be
\be\label{shiftagain}
S_{an} = 2i\kappa\xi \left(  2M-N_{an}\right),\qquad
 S_{an}^{\dagger}= -2i\kappa\partial_\xi \, .
 \ee
The shift operators do not commute with $G$, since
\be
[G_{an},S_{an}] = 2S_{an}\, , \qquad [G_{an},S^\dagger_{an}]=-2S^\dagger_{an}\, ,
\ee
and hence $S^\dagger$ is no longer the hermitian conjugate of $S$. In fact,
its hermitian conjugate is
$S^\ddagger =-S^\dagger$, and hence
\be\label{worldlineshort}
\{S_{an},S_{an}^\ddagger\} =  4\kappa^2 \left(N_{an} -2M\right).
\ee

 Again following the general procedure, we define  `improved' $ISU(1|1)$
 supersymmetry generators
\begin{equation}
{\tilde Q}_{an}=Q_{an}+ \frac{i}{2\kappa} S^{\dagger}_{an}
= z\partial_\zeta\, .
\end{equation}
As this operator commutes with $G_{an}$, we have
\begin{equation}
\tilde Q_{an}^\ddagger = \tilde Q_{an}^\dagger =Q_{an}^\dagger
- {\frac{i}{2\kappa}}S_{an}
= \zeta\partial_z\, .
\end{equation}
If we now define the new $U(1)$ generator
\be
\tilde C_{an}  = C_{an} +  \left(N_{an} -2M\right),
\ee
which yields precisely the same redefinition as in (\ref{ctilde}),
then the new $ISU(1|1)$ generators are
\begin{align}\label{pshorter}
P_{an} &= -i\partial_z \, ,\qquad P_{an}^\ddagger = 2i\kappa z\,, \nonumber \\
\Pi_{an} &= \partial_\zeta\, ,\qquad \Pi_{an}^\ddagger = 2\kappa\zeta\, ,\nonumber \\
\tilde Q_{an} &= z\partial_\zeta \, , \qquad
\tilde Q_{an}^\ddagger = \zeta\partial_z \, ,\nonumber\\
\tilde C_{an} &=z \partial_z + \zeta\partial_\zeta \, .
\end{align}
One may verify that these operators obey the (anti)commutation relations of $ISU(1|1)$.

As the `analytic' $ISU(1|1)$ generators now act on functions of $(z,\zeta)$ alone,
they evidently (anti)commute with $S_{an}$ and $S^\ddagger_{an}$, which act on functions
of $\xi$ alone. As a consequence, the variable $\xi$ can no longer be interpreted
as a Nambu-Goldstone variable for broken $ISU(1|1)$ supersymmetry, as it was
in \cite{Ivanov:2005vh}. Instead, it can be interpreted as a Nambu-Goldstone variable
for the symmetry generated by $S$. As we now explain, this is the generator of
a worldline supersymmetry, so the expansion of a wavefunction in $\xi$
is the  ($ISU(1|1)$-invariant)  expansion of a worldline superfield.

\subsection{Worldline supersymmetry revisited}

The anticommutation relation (\ref{worldlineshort}) implies that
\be\label{worldlinelong}
\{S,S^\ddagger\} = 2\kappa H_{susy} \, , \qquad
H_{susy} =  H -4\kappa M\,,
\ee
and one can similarly show that $\{S,S\}=0=\{S^\ddagger,S^\ddagger\}\,$.
It is therefore natural to interpret $S$ as an ${\cal N}{=}2$  worldline supersymmetry
charge, for Hamiltonian $H_{susy}\,$, but the assumption that $M\le0$ is crucial
to this interpretation because $S^\ddagger$ is otherwise not the hermitian
conjugate of $S$ with respect to a non-negative norm. Indeed,
the anticommutator (\ref{worldlineshort}) would, if valid, imply that
$N\ge 2M$, which would exclude states with  $N{<}2M$ when $M{>}0\,$. As noted earlier,
we must revert to the $G=1$ norm when $N{<}2M\,$,  in which case the
anticommutator  (\ref{worldlineshort}), and hence (\ref{worldlinelong}),
is modified. One finds that
\be\label{worldlinelong1}
\{S,S^\ddagger\} = 2\kappa |H_{susy}| \qquad (M{>}0)\,.
\ee
One could attempt to interpret this as a supersymmetry anticommutator
with $|H_{susy}|$ as a new Hamiltonian but this would be pointless as
it does not imply worldline supersymmetry of the planar superflag model.
For this reason, the planar superflag model has a hidden worldline supersymmetry
only for $M\le0$, so let us now assume that $M{\le0}\,$.

A standard consequence of \p{worldlinelong} is that $S$ and $S^\ddagger$ can only annihilate 
states that are annihilated by $H_{susy}$, which are eigenstates of $H$ with energy $4\kappa M\,$.
Given that $H_{an} = 2\kappa N_{an}$, for positive $\kappa$, and $M\leq0$, such states can exist
only when $M{=}0\,$, in which case they are zero-energy states.  In standard supersymmetric quantum mechanics, all zero-energy eigenstates must be annihilated by all supersymmetry charges. Our case is slightly different: a given zero-energy eigenstate need not be annihilated by $S$, or $S^\ddagger$, but if it is not then the resulting state has zero norm. This follows from the expressions \p{shiftagain} and \p{PanN}, and the fact that $\Psi^{(0)}_{an} = \psi^{(0)}_-$ has zero norm at $M=0\,$ (see \p{invnorm2}). Nevertheless, it is still true that all {\it physical} zero energy states are annihilated by both $S$ and $S^\ddagger$ because the physical subspace is spanned by equivalence classes of states modulo the addition of a zero-norm state. The number of these physical ground states is precisely half the total number of ground states, and hence non-zero, so the worldline supersymmetry is restored at $M{=}0\,$. This is expected from the equivalence
with the superplane model, for which we know that the worldline supersymmetry is unbroken.
In contrast, there are no supersymmetric ground states when $M{<}0\,$,
so  {\it worldline supersymmetry is spontaneously broken for $M{<}0\,$}.

These quantum results have classical analogs. To see this, we observe that
the charges $S$ and $S^\ddagger$ generate transformations of the phase-space variables
that  leave invariant the phase-space form of the classical action, which is given
in \cite{Ivanov:2005vh}.  After solving the phase-space constraints and eliminating
the momentum variables, one finds
the infinitesimal transformation laws
\bea\label{sfsusytrans}
\delta z &=& -\epsilon\xi\left(\dot z + \bar\xi \dot\zeta\right), \nonumber\\
\delta\zeta &=& =-\left[\left(1+\bar\xi\xi\right)\dot z + \bar\xi\dot\zeta\right]
\bar\epsilon\, ,\nonumber\\
\delta\xi &=& -2i\kappa\, \bar\epsilon\, ,
\eea
where $\epsilon$ is the complex anticommuting parameter, with complex conjugate  $\bar\epsilon$.
The transformations of $(\bar z,\bar\zeta,\bar\xi)$ are obtained by taking the complex conjugate.
One may verify that these transformations leave invariant the classical Lagrangian (\ref{planarSF}),
up to a total derivative, for any value of $M\,$, and have the same on-shell
closure $\sim 2\kappa \partial_t$ for all relevant variables. Note that the transformations of $(z,\zeta)$
are on-shell equivalent to those of (\ref{SUSYtran}), and that they are compatible with
the relation $\dot\zeta= -\xi\dot z$ since
$\delta(-\dot\zeta/\dot z) =  2\kappa i \, \epsilon$, on shell.

Although the on-shell relation $\dot\zeta= -\xi\dot z$ suggests that $\xi$ is a
`composite' variable at $\dot z \neq 0\,$,
it should now be recalled that it becomes an independent variable in a classical ground state
corresponding to $\dot z =0\,$, at least when $M{<}\,0$. Then its inhomogeneous transformation
implies that it is a Nambu-Goldstone variable for spontaneously broken worldline supersymmetry.
Thus, classical supersymmetry is spontaneously broken when $M{<}0\,$. The $M{=}0$ case
is different because, as mentioned earlier, $\xi$ can then be `gauged away' in a classical ground
state and the $\delta\xi$  transformation of \p{sfsusytrans} becomes just a particular case
of the corresponding gauge transformation. So classical worldline supersymmetry is unbroken when $M{=}0\,$. This of course could be anticipated from the equivalence of the $M{=}0$ planar superflag model to the superplane model. The classical physics therefore parallels the quantum physics:
supersymmetry is spontaneously broken for $M{<}0$ but restored at $M{=}0\,$.

\section{Conclusions}
\setcounter{equation}{0}

Earlier studies of super-Landau models \cite{Ivanov:2003qq,Ivanov:2004yw,Ivanov:2005vh}
concluded that these models have ghosts in all Landau levels but some number of
the low-lying ones, as might be expected for a theory with `higher-derivative' fermion kinetic
terms, but this conclusion was grounded on a particular choice of (super)Hilbert space norm.
Here we have investigated the possibility of other norms consistent with symmetries
and hermiticity of the Hamiltonian. We have found that there is an alternative
norm. This alternative norm is positive for the superplane Landau model, as is implicit
in the previous work of Hasebe \cite{Hasebe:2005cm}, and hence for the equivalent
$M=0$ planar superflag model. The alternative norm is also positive for the $M{<}0$ planar
superflag Landau models, while for $M{>}0$ the positive norm is  a `dynamical' combination
of both possibilities (in the sense that the choice depends on the level, and hence
on the Hamiltonian). Thus, it is always possible to find a positive norm. However, this positive
norm is not always  positive-definite because there are zero-norm states when $2M$
is a non-negative  integer\footnote{This phenomenon was shown in \cite{Ivanov:2005vh}
to be associated with a fermionic gauge-invariance.}. This means  that  the positive
norm cannot always be identified with the natural positive-definite norm in the basis
for which the Hamiltonian is diagonal, so merely noting the  existence of the latter
is not sufficient, in general,  to `exorcize'  the super-Landau ghosts; the detailed analysis
performed here was necessary.

The possibility of modifying the Hilbert space norm in order
to convert an apparently unphysical quantum theory into a physical one
is the underlying  theme of  `${\cal PT}$-symmetric' quantum theory, and
a number of methods for investigating these possibilities have been developed
in this context. Here we have taken over these methods, extending them to
models with anticommuting variables. Apart from the basic point that one may
be able to adjust the (super)Hilbert space norm, via a `metric operator'
so as to achieve a positive inner product, the main consequence of a redefined
norm is a redefined notion of hermitian conjugation. Specifically, operators
that do not commute with the metric operator have hermitian conjugates that do not coincide
with the naive conjugate. For the planar super-Landau models investigated here,
we found that the supersymmetry charges have hermitian conjugates that are
shifted, relative to their naive hermitian conjugates, by conserved
`shift' operators.

Remarkably, these `shift' operators are worldline supersymmetry  charges, analogous to those
noted by Hasebe in his version of the superplane Landau model  \cite{Hasebe:2005cm}.  Classically,
these charges generate transformations of the variables that leave the Lagrangian invariant,
up to a total derivative, and the classical ground state solution is supersymmetric;
this feature is maintained in the quantum theory, since the quantum ground state
is annihilated by the quantum supersymmetry charges.  Although the worldline
supersymmetry algebra is the standard one of supersymmetric quantum mechanics,
the form of the supersymmetry  transformations is non-standard. It would be
interesting to see whether there is a superspace version of the model that makes
manifest the invariance of the classical action.

The `hidden' worldline supersymmetry of planar super-Landau models emerges most naturally
in the more general planar superflag models when $M{<}0$ because the additional
anticommuting parameter of these models then has a natural interpretation as
the Nambu-Goldstone variable for a broken  worldline supersymmetry. Quantum mechanically,
the worldline supersymmetry is spontaneously broken because the supersymmetry charges
fail to annihilate the ground state. One might be tempted to conclude that the Witten
index is therefore zero and that, as a consequence, the worldline supersymmetry will
remain spontaneously broken for any (non-positive) value of the parameter $M$ (since
quantum  corrections to models of supersymmetric quantum mechanics generally raise
the energy of any state that would otherwise `accidentally'  have zero energy).
However, this conclusion would not be correct;  not because quantum corrections fail to raise
the energy of an otherwise zero-energy state but because the spectrum changes
discontinuously at $M{=}0$ due to the vanishing of the norm of half the lowest Landau
level states.  Thus,  worldline supersymmetry is restored at $M{=}0$ by a novel mechanism.

The discontinuity in the spectrum at $M{=}0$ suggests that the Witten index
is discontinuous too,  but the infinite degeneracy of the lowest Landau level
in planar Landau models may instead mean that the index is ill-defined. For this
reason, among others, it would be interesting  to know what happens for the
spherical super-Landau models, for which the degeneracies at each level are finite.
Of course, the issue arises only if the spherical super-Landau models also exhibit
a `hidden' worldline supersymmetry, and it  remains to be seen whether this
is the case.

\section*{Acknowledgements}

T.C. and L.M. thank the Institute for Advanced Study for its
hospitality and support, and for providing a stimulating environment in which
part of this work was completed. We thank  Carl Bender,  Shanta de Alwis, Jaume Gomis,
Jurek Lukierski,  Robert Myers,  Andrei Smilga,  Juan Maldacena
and Dima Sorokin for useful discussions.
This material is based upon work supported by the
National Science Foundation under Grants No's. 0303550 and 0555603. E.I.
acknowledges a partial support from the Grant RFBR 06-02-16684, RFBR-DFG Grant
06-02-04012 and the Grant INTAS-05-7928. P.K.T. thanks the EPSRC for
financial support.

\end{document}